\documentclass[12pt,aps,preprint,nofootinbib,superscriptaddress]{revtex4}

\usepackage{epsfig}
\usepackage{amssymb}
\usepackage{amsmath}
\usepackage{amsfonts}
\usepackage{graphics}
\usepackage{cancel}
\usepackage{bbm}
\usepackage{mathrsfs}
\usepackage{axodraw}

\newcommand{\Refs}{Refs.}
\newcommand{\Ref}{Ref.}

\newcommand{\eq}{Eq.}
\newcommand{\eqs}{Eqs.}
\newcommand{\fig}{Fig.}
\newcommand{\Fig}{Fig.}

\newcommand{\bea}{\begin{eqnarray}}
\newcommand{\eea}{\end{eqnarray}}
\newcommand{\type}[1]{type-#1}

\newcommand{\be}{\begin{equation}}
\newcommand{\ee}{\end{equation}}

\newcommand{\ba}{\begin{array}}
\newcommand{\ea}{\end{array}}

\newcommand{\ie}{\emph{i.e.}}
\newcommand{\eg}{\emph{e.g.}}
\newcommand{\cf}{\emph{c.f.}}
\newcommand{\CP}{$C\!P$~}
\newcommand{\hc}{\mathrm{H.c.}}

\DeclareMathOperator{\diag}{diag}   

\allowdisplaybreaks

\begin{document}
\title{Non-unitary Leptonic Mixing and Leptogenesis}

\author{Stefan Antusch}
\email[]{antusch@mppmu.mpg.de} \affiliation{Max-Planck-Institut f\"ur Physik (Werner-Heisenberg-Institut), F\"ohringer Ring 6, 80805 M\"unchen, Germany}

\author{Steve Blanchet}
\email[]{sblanche@umd.edu} \affiliation{Maryland Center for Fundamental Physics, University of Maryland, College Park, MD, 20742, USA}

\author{Mattias~Blennow}
\email[]{blennow@mppmu.mpg.de} \affiliation{Max-Planck-Institut f\"ur Physik (Werner-Heisenberg-Institut), F\"ohringer Ring 6, 80805 M\"unchen, Germany}

\author{Enrique Fernandez-Martinez}
\email[]{enfmarti@mppmu.mpg.de} \affiliation{Max-Planck-Institut f\"ur Physik (Werner-Heisenberg-Institut), F\"ohringer Ring 6, 80805 M\"unchen, Germany}

\date{\today}

\begin{abstract}
We investigate the relation between non-unitarity of the leptonic mixing matrix and leptogenesis. We discuss how all parameters of the canonical \type{I} seesaw mechanism can, in principle,
be reconstructed from the neutrino mass matrix and the deviation of the effective low-energy leptonic mixing matrix from unitary. When the mass $M'$ of the lightest right-handed
neutrino is much lighter than the masses of the others, we show that its decay asymmetries within flavour-dependent leptogenesis can be expressed in terms of two contributions, one depending on
the unique dimension five ($d=5$) operator generating neutrino masses and one depending on the dimension six ($d=6$) operator associated with non-unitarity. In low-energy seesaw scenarios where small lepton
number violation explains the smallness of neutrino masses, the lepton number conserving $d=6$ operator contribution generically dominates over the $d=5$ operator contribution
which results in a strong enhancement of the flavour-dependent decay asymmetries without any resonance effects. To calculate the produced final baryon asymmetry, the flavour equilibration
effects directly related to non-unitarity have to be taken into account. In a simple realization of this \emph{non-unitarity driven leptogenesis}, the lower bound on $M'$ is found to be about $10^8$~GeV at the onset of the strong washout regime, more than one order of magnitude below the bound in ``standard'' thermal leptogenesis.
\end{abstract}


\preprint{MPP-2009-175} \preprint{UMD-PP-09-056}

\maketitle

\section{Introduction}
Non-unitarity of the leptonic mixing matrix at low energies is a generic manifestation of new physics in the lepton sector often related to the mechanism responsible for the generation
of neutrino masses. Non-unitarity appears whenever additional heavy particles mix with the light neutrinos or their charged lepton partners. After integrating the heavy states out of the
theory, the $3 \times 3$ submatrix of the light neutrinos remains as an effective mixing matrix. This low-energy leptonic mixing matrix is, in general, not unitary.

One example where non-unitarity is predicted is the generic \type{I} seesaw mechanism~\cite{Minkowski:1977sc,Mohapatra:1979ia,Yanagida:1979as,GellMann:1980vs}, where the Standard Model (SM)
is extended by (typically three) right-handed neutrinos. If the \type{I} seesaw mechanism operates at energies as high as the Grand Unification scale (GUT scale), then non-unitarity effects
are tiny. However, if the seesaw mechanism is realized at low energies close to the electroweak scale, then non-unitarity is enhanced and can be observable. It may then  provide important
hints to the origin of neutrino masses. It is important to note that, in low-energy seesaw scenarios, the smallness of the neutrino masses is not explained by the largeness of the seesaw scale.
However it can be explained in a technically natural way by a lepton number symmetry which is broken only by a small amount \cite{Branco:1988ex}.

While the neutrino masses in the \type{I} seesaw mechanism are effectively described by the unique lepton number violating Weinberg operator of $d=5$, the non-unitarity of the leptonic
mixing matrix is generated by the lepton number conserving $d=6$ operator contributing to the kinetic terms of the neutrinos \cite{DeGouvea:2001mz,Broncano:2002rw}. In low-energy seesaw
scenarios with approximately conserved lepton number, the $d=6$ operator can cause significant effects, since it is not suppressed by the smallness of the neutrino masses.

One attractive feature of the seesaw mechanism is that it can explain the observed baryon asymmetry of the Universe via the mechanism of leptogenesis~\cite{Fukugita:1986hr} (for a recent
review, see \Ref~\cite{Davidson:2008bu}). If the seesaw mechanism, and thus also the mechanism of thermal leptogenesis, operates at high energies, the decay asymmetries for leptogenesis are typically dominated
by the $d=5$ operator. Unfortunately the high-energy parameters which control leptogenesis cannot be fully reconstructed from the measurements at low energy, since
combinations of the Yukawa couplings different from those in the $d=5$ operator also appear.

On the other hand, if the seesaw mechanism operates at lower energies, predicting an observable non-unitarity of the leptonic mixing matrix, one may in principle obtain enough information from
low-energy measurements to reconstruct the full Lagrangian and, therefore, the parameters that control leptogenesis \cite{Broncano:2002rw,Broncano:2003fq,Xing:2009vb}.

The purpose of this paper is to clarify the relation between the high-energy parameters that control successful leptogenesis and their low-energy manifestations, \ie, neutrino masses and
mixings and deviations from unitarity of the leptonic mixing matrix, making special emphasis on the latter. In Section \ref{sec:reconstruction} we describe in detail the method and conditions
under which the full high-energy Lagrangian can be reconstructed from the low-energy effects. In Section \ref{sec:model} we introduce a ``minimalistic'' low-scale seesaw model with three
right-handed neutrinos in which the smallness of neutrino masses is explained by an approximate lepton number symmetry. However,
deviations from unitary mixing induced by the $d=6$ operator are not protected by the symmetry and can be sizeable, leading
to effects that could, in principle, be tested in precision electroweak measurements. In Section \ref{sec:lepto} we discuss
how, in this model, leptogenesis could be driven by the $d=6$ operator that induces the deviation from unitary mixing via flavoured leptogenesis. In Section \ref{sec:equilibration} we point out that the same $d=6$ operator that drives flavoured leptogenesis can also lead to a flavour equilibration, which could wash out the generation of lepton
number. We also discuss the conditions under which successful leptogenesis can occur. Finally, in Section \ref{sec:conclusions}, we summarize and discuss our results.

\section{Non-unitarity relation to high-energy observables in the \type{I} seesaw model}
\label{sec:reconstruction} 

In this section we describe how the full Lagrangian of the \type{I} seesaw can, in principle, be reconstructed from the low-energy observations of neutrino masses and mixings, including deviations from unitary mixing. The conditions under which this reconstruction is possible were described in \Ref~\cite{Broncano:2002rw}, while a method to realize the reconstruction was
outlined in \Ref~\cite{Broncano:2003fq}. Here we present a new algorithm to perform the reconstruction through which the high-energy parameters can be derived more easily.

Let us consider the Lagrangian of the standard \type{I} seesaw model which consists of the one for the SM plus an extra piece containing the allowed couplings between the SM fields and
additional gauge singlet fermions (\ie, right-handed neutrinos)
 $N_\mathrm{R}^i$:
\begin{eqnarray}\label{eq:The3FormsOfNuMassOp}
\mathscr{L} &=& \mathscr{L}_\mathrm{SM} -\frac{1}{2} \overline{N_\mathrm{R}^i} M^N_{ij} N^{c j}_\mathrm{R} -(Y_{N})_{i\alpha}\overline{N_\mathrm{R}^i} \widetilde \phi^\dagger
\ell^\alpha_\mathrm{L} +\hc\; .
\end{eqnarray}
Here, $\phi$ denotes the SM Higgs field, which breaks the electroweak (EW) symmetry after acquiring its vacuum expectation value (vev) $v_{\mathrm{EW}}$, and we have used the definition
$\tilde \phi = i \tau_2 \phi^*$.

The low-energy effects of the three-family low-scale seesaw model, from the point of view of neutrino oscillation experiments, is given by two effective operators, one of mass dimension five
and one of mass dimension six. 
The $d=5$ operator is the ubiquitous lepton number violating Weinberg operator
\begin{eqnarray}
\delta{\cal L}^{d=5} = \frac{1}{2}\, c_{\alpha \beta}^{d=5} \,\left( \overline{L^c}_{\alpha} \tilde \phi^* \right) \left( \tilde \phi^\dagger \, L_{ \beta} \right) + \hc \;,
\end{eqnarray}
which is the lowest-dimensional effective operator for generating neutrino masses using the field content of the SM. The coefficient matrix $c_{\alpha \beta}^{d=5}$ is
\begin{eqnarray}
c_{\alpha \beta}^{d=5} = -(Y_N^T)_{\alpha i} (M_N)_{ij}^{-1} (Y_N)_{j\beta}
\end{eqnarray}
and relates to the low-energy neutrino mass matrix as
\begin{eqnarray}\label{Eq:d5AndMnu}
m_\nu = v^2_{\mathrm{EW}} c_{}^{d=5}\;.
\end{eqnarray}
The effective $d=6$ operator
\begin{eqnarray}
\delta{\cal L}^{d=6} = c^{d=6}_{\alpha \beta} \, \left( \overline{L}_{\alpha} \tilde \phi \right) i \cancel{\partial} \left(\tilde \phi^\dagger L_{ \beta} \right)
\end{eqnarray}
conserves lepton number and, after EW symmetry breaking, contributes to the kinetic terms of the neutrinos. After their canonical normalization, they generate a non-unitary leptonic mixing
matrix $N$ as well as non-universal couplings proportional to $N^\dagger N$ of the neutrinos to the $Z$ boson (see, \eg, \Refs~\cite{DeGouvea:2001mz,Broncano:2002rw,Antusch:2006vwa}). The
coefficient matrix $c_{\alpha \beta}^{d=6}$ is given by (see, \eg, \Ref~\cite{Broncano:2002rw})
\begin{eqnarray}
c_{\alpha \beta}^{d=6} = \sum_i (Y_N^\dagger)_{\alpha i} (M_N)_{ii}^{-2} (Y_N)_{i\beta}\;,
\end{eqnarray}
in the basis where $M_N$ is diagonal.

If we parametrize the non-unitary leptonic mixing matrix $N$ as \cite{FernandezMartinez:2007ms}
\begin{equation}
N = (1 + \eta)\,U\;, \label{param}
\end{equation}
where $\eta$ is Hermitian and $U$ is unitary, then $\eta_{\alpha \beta}$ is related to the coefficient matrix $c_{\alpha \beta}^{d=6}$ by
\begin{eqnarray}\label{Eq:EpsAndNonU}
\eta_{\alpha \beta} = - v^2_{\mathrm{EW}} c_{\alpha \beta}^{d=6}/2 \; .
\end{eqnarray}
In \Refs~\cite{Langacker:1988ur,Nardi:1994iv,Tommasini:1995ii,Antusch:2006vwa,Antusch:2008tz}, the following constraints on these parameters at the $90~\%$ C.L.\ were derived:
$\eta_{ee}<2.0 \cdot 10^{-3}$, $\eta_{e \mu}<5.9 \cdot 10^{-5}$, $\eta_{e \tau}<1.6 \cdot 10^{-3}$, $\eta_{\mu \mu}<8.2 \cdot 10^{-4}$, $\eta_{\mu \tau}<1.0 \cdot 10^{-3}$ and $\eta_{\tau
\tau}<2.6 \cdot 10^{-3}$.

In the \type{I} seesaw the full $6\times 6$ mixing matrix $U_{\rm tot}$ is the unitary matrix that diagonalizes the extended neutrino mass matrix: %
\begin{equation}
U_{\rm tot}^T \left(
\begin{array}{cc}
0 & m_D^T \\ m_D & M_N
\end{array}
\right) U_{\rm tot} = \left(
\begin{array}{cc}
m & 0 \\ 0 & M
\end{array}
\right), \label{diag}
\end{equation}
where $m_D = v_{EW} Y_N$ and $M_N$ are the neutrino's Dirac and Majorana mass matrices and $m$ and $M$ are diagonal matrices. It is easier to perform the diagonalization in two steps: first
a block-diagonalization and then two unitary rotations to diagonalize the mass matrices of the light and heavy neutrinos, \ie, %
\begin{equation}
U_{\rm tot} = \left(
\begin{array}{cc}
A & B \\ C & D
\end{array}
\right) \left(
\begin{array}{cc}
U & 0 \\ 0 & V
\end{array}
\right), \label{block}
\end{equation}
where $U$ and $V$ are unitary matrices. Without loss of generality, we can choose a basis for the heavy singlets such that $V=I$ and $M_N = M$. When performing the block diagonalization,
the mixing between the light and heavy neutrinos is suppressed when $M_N > m_D$ so that
\begin{equation}
 B \simeq \Theta = m_D^\dagger M_N^{-1}.
 \label{theta}
\end{equation}

We can exploit the suppression of \eq~(\ref{theta}) to write the unitary block diagonalization as the exponential expansion of an anti-Hermitian matrix:
\begin{equation}
\left(
\begin{array}{cc}
A & B \\ C & D
\end{array}
\right) = \exp \left(
\begin{array}{cc}
0 & \Theta \\ -\Theta^\dagger & 0
\end{array}
\right) = \left(
\begin{array}{cc}
1-\frac{1}{2}\Theta \Theta^\dagger & \Theta \\ -\Theta^\dagger & 1-\frac{1}{2}\Theta^\dagger \Theta
\end{array}
\right) + {\cal O}(\Theta^3).
\end{equation}
The block-diagonalization yields the complex symmetric neutrino mass matrix
\begin{equation}
 m_\nu = - m_D^T M_N^{-1} m_D,
\label{diagon}
\end{equation}
which can be diagonalized by a unitary transformation $U$ such that $m = \diag(m_1,m_2,m_3) = U^T m_\nu U$. Notice that the mixing matrix of the three light neutrinos is given by
\begin{equation}
N = AU  = (1 + \eta)U = (1 - \Theta \Theta^\dagger/2) U,
\end{equation}
As described above, $\eta = -\Theta \Theta^\dagger/2$ exactly contains the coefficients $c^{d=6}$ of the $d=6$ operator. In particular, note that this implies that $c^{d=6}$ is Hermitian and
positive semidefinite.

Assuming that the low-energy observables $U$, $m$ and $\eta$ have been measured, it is then natural to ask the question of whether one can reconstruct the high-energy parameters contained in
$m_D$ and $M$. In order to split the low- and high-energy observables, we make use of the parametrization proposed in \Ref~\cite{Casas:2001sr} for the Dirac mass matrix $m_D$, which is
introduced as follows: From \eq~(\ref{diagon}) we have that $-\sqrt{m^{-1}} U^T m_D^T M^{-1} m_D U \sqrt{m^{-1}} = 1$. Thus, defining $R \equiv i \sqrt{M^{-1}} m_D U \sqrt{m^{-1}}$, we have the
condition $R^TR = 1$. We will consider here the case in which the number of heavy right-handed singlets is equal to the number of light neutrinos, and thus, the matrix $R$ must be a complex
orthogonal matrix. Multiplying by inverses of matrices, the definition of $R$ can be rewritten as
\begin{equation}
 m_D = - i \sqrt{M} R \sqrt{m} U^\dagger
\label{ciparam}
\end{equation}
Notice that the matrix $U$ that appears in the parametrization of \eq~(\ref{ciparam}) is not the neutrino mixing matrix
that describes the neutrino couplings in charged-current (CC) interactions $N=(1+\eta)U$, but only the unitary part of the CC mixing diagonalizing $m_\nu$. In \eq~(\ref{ciparam}) the observables of the $d=5$ operator describing the light neutrino mass matrix are all
contained within $U$ and $m$, while $R$ and $M$ contain the missing information in order to reconstruct the high-energy parameters. In order to perform the reconstruction we will assume that
both the $d=5$ operator, \ie,  $U$ and $m$, and the $d=6$ operator $\eta$ are known. Notice that the deviations from unitary mixing encoded in $\eta$ can be probed and potentially measured
through electroweak decays \cite{Langacker:1988ur,Nardi:1994iv,Tommasini:1995ii,Antusch:2006vwa,Abada:2007ux} as well as neutrino oscillation experiments
\cite{Bekman:2002zk,FernandezMartinez:2007ms,Holeczek:2007kk,Goswami:2008mi,Altarelli:2008yr,Antusch:2009pm}.

Using the parametrization of \eq~(\ref{ciparam}) in the expression for the $d=6$ operator  $2 \eta = - m^\dagger_D M^{-2} m_D$, we define the matrix $H$ as
\begin{equation}
 H \equiv - \sqrt{m^{-1}} U^\dagger 2 \eta U \sqrt{m^{-1}} = (R^*)^{-1} M^{-1} R.
\label{condi}
\end{equation}
With this definition, $H$ is a positive semidefinite Hermitian matrix, since it can be decomposed as $F F^\dagger$, and contains all of the available low-energy information. Equation~(\ref{condi}) is known
as the conjugate diagonalization (or simply the ``condiagonalization'') of the matrix $H$. Notice that replacing $(R^*)^{-1}$ by $R^{-1}$ in \eq~(\ref{condi}) would reduce it to a normal
diagonalization. It can be shown that all Hermitian positive definite matrices can be condiagonalized and the solution is unique under the requirement that the $M_i$ are real and positive.
The simplest way of reconstructing $R$ and $M$ from the matrix $H$ is to note that $H^* H = H^T H$ is a complex symmetric matrix and that
\begin{equation}
 H^* H = R^{-1} M^{-1} R^* (R^*)^{-1} M^{-1} R = R^T M^{-2} R,
\end{equation}
and thus, $R$ is the complex orthogonal matrix which diagonalizes $H^* H$ with corresponding eigenvalues $M_i^{-2}$. It should be noted that the reconstruction of $R$ and $M$ has been
previously studied in \Ref~\cite{Broncano:2003fq}. However, the simple reconstruction algorithm presented here, only involving the diagonalization of a complex symmetric matrix, is new.
Notice that, for the algorithm to work and for \eq~(\ref{condi}) to have solutions, it is vital that the matrix $H$ is a positive definite matrix. This will be guaranteed as long as the $d=6$
operator $c^{d=6}$ is also positive definite. By construction $c^{d=6}$ is positive semidefinite so the only case in which $H$ is not positive definite is when $c^{d=6}$ has one or more zero
eigenvalues. This could happen for two reasons: either the number of heavy right-handed neutrinos is smaller than the number of light neutrinos, or the rows of the Yukawa matrix are not
linearly independent. In any of these cases the number of parameters in the high-energy theory is smaller than in the full case considered here and the reconstruction of the Lagrangian would
then be easier, sometimes without the need of involving the $d=6$ operator but only through $U$ and $m$ (see, \eg, \Ref~\cite{Gavela:2009cd}). If, on the other hand, the number of heavy
right-handed singlets is larger than the number of light neutrinos, then the information encoded in the $d=6$ operator $\eta$ plus the neutrino masses and mixings $m$ and $U$ is not sufficient to
reconstruct the full high-energy Lagrangian. We find these limitations to be in agreement with \Ref~\cite{Broncano:2003fq}, where the reconstruction algorithm relies on $\eta$ being
invertible.

\section{The three-family low scale seesaw scenario}
\label{sec:model}

The ``minimalistic'' low-scale seesaw model we present here is a \type{I} seesaw model with three right-handed neutrinos to which we additionally impose a softly broken ``lepton number''-like
(global) U(1) symmetry where the charge of the SU(2)$_{\mathrm{L}}$ doublets $\ell^f_\mathrm{L}$ is opposite to that of the field $N_\mathrm{R}^3$ but equal to that of the field
$N_\mathrm{R}^2$. We assign zero lepton number to $N_\mathrm{R}^1$. In the symmetry limit, $M^N_{ij}$ and $Y_{N}$ are forced to have the form:
\begin{eqnarray}
Y_{N} = \begin{pmatrix}
 0 & 0 & 0 \\
y_e & y_\mu & y_\tau \\
0 & 0 & 0
\end{pmatrix} \; , \quad
M^N= \begin{pmatrix}
M' & 0 & 0 \\
0 & 0 & M \\
0 & M & 0  \end{pmatrix} \; .
\end{eqnarray}
At this level, the neutrinos are exactly massless, but non-unitarity of the leptonic mixing matrix is already induced. When small soft breaking terms $\mu_i$ and $\mu'_i$ are allowed, this
rigid structure is perturbed to
\begin{eqnarray}
Y_{N} = \begin{pmatrix}
\mu'_e & \mu'_\mu & \mu'_\tau \\
 y_e & y_\mu & y_\tau \\
  \mu_e & \mu_\mu & \mu_\tau
\end{pmatrix} \; , \quad
M^N= \begin{pmatrix}
M' & \mu'_4 & \mu'_5 \\
\mu'_4 & \mu_4 & M \\
 \mu'_5 & M & \mu_5 \end{pmatrix} \; ,
\end{eqnarray}
and masses for the light neutrinos, suppressed by the small $\mu_i$ and $\mu'_i$ parameters, are generated. Notice that the $d=6$ operator is not protected by the lepton number symmetry and
would have a large leading order contribution
\begin{eqnarray}\label{d6op}
c_{\alpha \beta}^{d=6} = \frac{y_\alpha^* y_\beta}{M^2} + \ldots,
\end{eqnarray}
where the dots denote sub-leading terms proportional to $\mu_\alpha$ (or $\mu_4/M,\mu_5/M$) and $\mu'_\alpha$ (or $\mu'_4/M,\mu'_5/M$). In particular, the six moduli $|\eta_{\alpha\beta}|$
depend only on the three parameters $|y_e/M|,|y_\mu/M|,|y_\tau/M|$. This implies that at leading order the rank of the $d=6$ operator is one and that the reconstruction algorithm presented in
Section~\ref{sec:reconstruction} would not be applicable. On the other hand, the number of parameters contributing to the $d=5$ and $d=6$ operator at leading order in this model is very
limited and they can actually be reconstructed with information mainly from $m$ and $U$. Indeed, the $d=5$ operator is given by \cite{Gavela:2009cd}:
\begin{eqnarray}\label{d5op}
c_{\alpha \beta}^{d=5} = (\mu_\alpha - \frac{\mu_5}{M} y_\alpha) \frac{1}{M} y_\beta + y_\alpha \frac{1}{M} (\mu_\beta - \frac{\mu_5}{M} y_\beta) + ...\;.
\end{eqnarray}
Following \Ref~\cite{Gavela:2009cd} both vectors $y_\alpha$ and $\mu_\alpha - (\mu_5/M) y_\alpha$ can be reconstructed from $m$ and $U$ up to an overall normalization. In particular,
\begin{eqnarray}
y_\alpha &\propto& \sqrt{1+\rho} U^*_{\alpha 3} + \sqrt{1-\rho} U^*_{\alpha 2}, \\ \rho &=& \frac{\sqrt{1+r} - \sqrt{r}}{\sqrt{1+r} + \sqrt{r}}
\end{eqnarray}
for normal hierarchy and
\begin{eqnarray}
y_\alpha &\propto& \sqrt{1+\rho}  U^*_{\alpha 2} + \sqrt{1-\rho} U^*_{\alpha 1}, \\ \rho &=& \frac{\sqrt{1+r} - 1}{\sqrt{1+r} + 1}
\end{eqnarray}
for inverted hierarchy. Here, $r=|\Delta m^2_{21}|/|\Delta m^2_{31}|$. If terms beyond leading order are considered for the $d=6$ operator, then the reconstruction described in
Sec.~\ref{sec:reconstruction} can be applied. In principle, this would also allow the extraction of the first Yukawa row $Y_{\alpha 1}=\mu'_{\alpha}$ if the
$d=6$ operator is known with sufficient precision.

\section{Leptogenesis in the three-family low-scale seesaw scenario}
\label{sec:lepto}

\subsection{Flavour-dependent decay asymmetries}

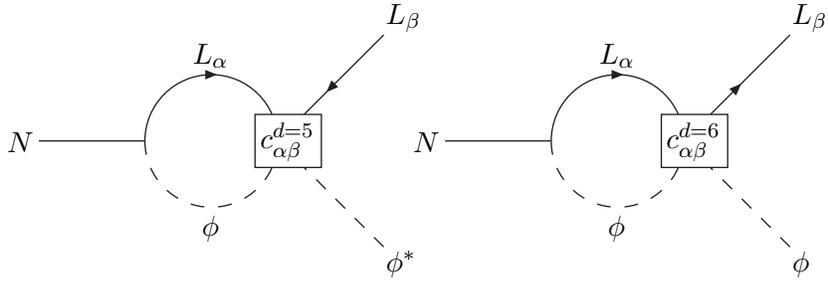
\begin{figure}
 \begin{center}
 \begin{picture}(150,90)(0,30)
  \Text(8,75)[r]{$N$}
  \Line(10,75)(50,75)
  \Text(75,103)[b]{$L_\alpha$}
  \ArrowArcn(75,75)(25,180,0)
  \Text(75,47)[t]{$\phi$}
  \DashCArc(75,75)(25,180,360){5}
  \Text(141.4,116.4)[bl]{$\bar L_\beta$}
  \ArrowLine(140,115)(100,75)
  \Text(141.4,33.6)[tl]{$\phi^*$}
  \DashLine(100,75)(140,35){5}
  \BBoxc(104,75)(25,20)
  \Text(104,75)[c]{$c^{d=5}_{\alpha\beta}$}
 \end{picture}
 \begin{picture}(150,90)(0,30)
  \Text(8,75)[r]{$N$}
  \Line(10,75)(50,75)
  \Text(75,103)[b]{$L_\alpha$}
  \ArrowArcn(75,75)(25,180,0)
  \Text(75,47)[t]{$\phi$}
  \DashCArc(75,75)(25,180,360){5}
  \Text(141.4,116.4)[bl]{$L_\beta$}
  \ArrowLine(100,75)(140,115)
  \Text(141.4,33.6)[tl]{$\phi$}
  \DashLine(100,75)(140,35){5}
  \BBoxc(104,75)(25,20)
  \Text(104,75)[c]{$c^{d=6}_{\alpha\beta}$}
 \end{picture}
 \end{center}
 \caption{\label{fig:diagrams}Effective operator decomposition of the diagrams leading to leptogenesis.}
\end{figure}

The flavour-dependent \CP asymmetries for the lightest right-handed neutrino (with mass $M'$) in the lepton flavour $\alpha$ are given by
\begin{eqnarray}
\varepsilon_{1,\alpha} &\simeq& \frac{1}{8\pi (Y Y^\dagger)_{11}} \sum_{j \not= 1} \left\{ \mbox{Im} \left[ Y^T_{\alpha 1} Y^\dagger_{\alpha j} (Y Y^\dagger)_{1 j} \frac{3}{2 \sqrt{x_j}} +
Y^T_{\alpha 1} Y^\dagger_{\alpha j} (Y Y^\dagger)_{j 1} \frac{1}{x_j} \right] \right\},
\end{eqnarray}
where $x_j = M_j^2/M_1^2 \gg 1$ was assumed (with $j=2,3$, $M_1 \equiv M'$, and with $M_2$, $M_3$ being the masses of $N_\mathrm{R}^2$ and $N_\mathrm{R}^3$). It is then possible to rewrite
$\varepsilon_{1,\alpha}$ as the sum of two contributions, that of the $d=5$ operator and that of the $d=6$ operator (see \Fig~\ref{fig:diagrams}):
\begin{eqnarray}\label{Eq:AssFromd=6}
\varepsilon_{1,\alpha} &\simeq& \frac{1}{8\pi (Y Y^\dagger)_{11}} \sum_{\beta} \left\{ \mbox{Im} \left[-\frac{3 M'}{2} Y_{1 \alpha }^*  \:c^{d=5}_{\alpha \beta} \:Y^\dagger_{\beta 1} + M'^2
\,Y_{1 \alpha}\:  c^{d=6}_{\alpha \beta}  \: Y^\dagger_{\beta 1} \right] \right\} \nonumber \\ &\simeq& \frac{M'^2}{8\pi (Y Y^\dagger)_{11}} \sum_{\beta} \left\{ \mbox{Im} \left[Y_{1
\alpha}\:  c^{d=6}_{\alpha \beta}  \: Y^\dagger_{\beta 1} \right] \right\},
\end{eqnarray}
where, in the last step, we have neglected the contribution from the $d=5$ operator, since it is protected by the lepton number symmetry that explains the smallness of neutrino masses, which is not the case for the $d=6$ operator.
Assuming the presence of an order one phase contribution to the imaginary part in Eq.~(\ref{Eq:AssFromd=6}), as well as $|Y_{1e}|\sim |Y_{1\mu}| \sim |Y_{1\tau}|$, we can further simplify
this expression and express it in terms of the non-unitarity parameters in Eq.~(\ref{Eq:EpsAndNonU}):
\begin{eqnarray}\label{CPasym}
\varepsilon_{1,\alpha} &\simeq& \frac{M'^2}{24\pi}  \sum_{\beta\neq \alpha}  c^{d=6}_{\alpha \beta} \; = \;  - \frac{1}{12\pi} \frac{M'^2}{v^2_{\mathrm{EW}}}\sum_{\beta\neq \alpha}
\eta_{\alpha \beta}\;.
\end{eqnarray}
For $M'=1$ TeV and $\eta_{\alpha \beta}$ of order $10^{-4}$ we obtain a large \CP asymmetry $\varepsilon_{1,\alpha}$ of order $10^{-3}$ without resorting to the usual enhancement for
quasi-degenerate masses of the heavy neutrinos~\cite{Covi:1996wh,Pilaftsis:1997jf}. As the contribution to $\varepsilon_{1,\alpha}$ originates from the $d=6$ operator (see also \Ref~\cite{GonzalezGarcia:2009qd}), we will refer to this scenario as \emph{non-unitarity driven leptogenesis}. 

It is crucial for our scenario that leptogenesis occurs when flavour effects are relevant~\cite{Abada:2006fw,Nardi:2006fx}, \ie, for a right-handed neutrino mass $M'<10^{12}$~GeV,
so that the \emph{flavoured} \CP asymmetries need to be considered instead of the total one $\varepsilon_{1}=\sum_\alpha \varepsilon_{1,\alpha}$, which is suppressed by the neutrino mass operator
($d=5$). As we will see next, another requirement for a non-zero baryon asymmetry is that the washout in each flavour is different. This is therefore an example of purely flavoured
leptogenesis~\cite{Nardi:2006fx,AristizabalSierra:2009bh}, where the asymmetry is generated exclusively due to flavour effects.

\subsection{Final baryon asymmetry}

An important parameter for flavoured leptogenesis is given by the decay (or washout) parameter, $K_{i\alpha}$, induced by the RH neutrino $N_{i}$, defined as the decay width over the Hubble
expansion rate when $T=M_i$:
\begin{equation}
K_{i\alpha}\equiv \frac{\Gamma_D (N_{i}\to \ell_{\alpha} \phi +\bar{\ell_{\alpha}}\phi^{\dagger})}{ H(T=M_i)}=\frac{|Y_{i\alpha}|^2 v_{\rm EW}^2}{m_{\star} M_i},
\end{equation}
where $m_{\star}\simeq 1.08\times 10^{-3}$~eV \cite{Buchmuller:2004nz}. For future use, we also define $K_i\equiv \sum_{\alpha}K_{i\alpha}$.

With the assumption $M' \ll M$ and $M'<10^9$~GeV, the asymmetry is generated by $N_1$ in the three-flavour regime~\cite{Abada:2006fw,Nardi:2006fx}. The contributions from the heavier states
are exponentially washed out in all flavours, as long as $K_{1\alpha}\gtrsim 3,~\forall \alpha \in \{e,\mu,\tau\}$~\cite{Blanchet:2008pw}. Moreover, as pointed out above, since the \CP asymmetry is non-zero due to a pure
flavour effect, it is crucial for the generation of the asymmetry that the washout is different in each flavour. Therefore, for simplicity, we will assume
 that the asymmetry is dominantly generated in one flavour, \ie, $\eta_{B\alpha}\simeq \eta_B$. The flavoured final asymmetry can be expressed as
\begin{equation}\label{finalasym}
\eta_{B\alpha}=0.88\times 10^{-2} \varepsilon_{1,\alpha} \;\kappa(K_{1\alpha}),
\end{equation}
where it was assumed that sphalerons decouple after the electroweak phase transition~\cite{Khlebnikov:1988sr,Harvey:1990qw}. For $K_{1\alpha}\gtrsim 3$, the efficiency factor $\kappa$ can be
approximated as~\cite{Giudice:2003jh}
\begin{equation}\label{washout}
\kappa(K_{1\alpha})\simeq \frac{0.5}{K_{1\alpha}^{1.16}}.
\end{equation}
In order to focus on the more relevant non-unitarity parameters and the mass scale $M'$, we will for now fix a washout $K_{1\alpha}= 5$, which is a typical value
 in the strong washout regime. This requires Yukawa couplings $|Y_{1
\alpha}|\sim 10^{-6}$, as needed for the TeV-scale seesaw mechanism. We further assume the value $K_1\equiv \sum_{\gamma}K_{1\gamma}=45$, such that $K_{1\alpha}\ll
K_{1\beta\neq\alpha}$. In other words, in terms of Yukawa couplings, we have the relation $2\,|Y_{1\alpha}|= |Y_{1\beta\neq \alpha}|$. Using \eqs~(\ref{Eq:AssFromd=6}), (\ref{finalasym}) and
(\ref{washout}), we have
\begin{equation}\label{finalasym2}
\eta_{B\alpha}\simeq -\frac{0.88\times 10^{-2}}{36\pi \,5^{1.16}} \frac{M'^2}{v_{\rm EW}^2}\sum_{\beta\neq \alpha}\eta_{\alpha \beta}\simeq 0.6\times 10^{-5}
\left(\frac{M'}{M}\right)^2\sum_{\beta\neq \alpha}y^*_\alpha y_\beta.
\end{equation}
This prediction should be compared to the measured value $\eta_B^{\rm CMB}=(6.2\pm 0.15)\times 10^{-10}$~\cite{Komatsu:2008hk}. With Yukawa couplings $y_{\beta}\sim \mathcal O(1)$, it is easy
to see that leptogenesis is possible with a mild hierarchy $M'/M\sim 10^{-2}$. Moreover, it is interesting to see that the scale of leptogenesis can be lowered, at least in principle, to the
weak scale. This is possible in our scenario because the purely flavoured contribution to the \CP asymmetry is not suppressed by the neutrino mass operator as in the usual case. The lower
bound on the scale of leptogenesis~\cite{Davidson:2002qv,Buchmuller:2002rq,Blanchet:2006be,Antusch:2006gy}, given by $M_1>3\times 10^9$~GeV at the onset of
 the strong washout, therefore does not apply. This was already noticed in \Ref~\cite{Blanchet:2008pw},
where, using the Casas--Ibarra parametrization~\cite{Casas:2001sr} in the limit $|\omega_{32}|\gg 1$, it was shown that the scale of leptogenesis could be lowered; the inverse seesaw model can
be shown to correspond to the extreme case $|\omega_{32}|\gg 100$. The seesaw model under consideration could therefore potentially offer an alternative to quasi-degenerate RH
neutrinos~\cite{Covi:1996wh,Pilaftsis:1997jf} to evade the gravitino bounds~\cite{Khlopov:1984pf,Ellis:1995mr,Moroi:1993mb,Pradler:2006hh}.

\section{Flavour equilibration}
\label{sec:equilibration} 

{F}rom the previous discussion it seems that the scale of leptogenesis could be lowered to the weak scale without any problem. However, an important effect was neglected, namely flavour
equilibration~\cite{AristizabalSierra:2009mq}. If flavours equilibrate, the final baryon asymmetry is proportional to the total \CP asymmetry
$\varepsilon_1=\sum_{\alpha}\varepsilon_{1\alpha}$, which is suppressed by the $d=5$ neutrino mass operator. We would thus recover the standard scenario, and the usual lower bound would
apply.

Let us now estimate how efficient flavour equilibration is in our case. The main processes are $\Delta L=0$ scatterings with off-shell $N_2$ and $N_3$, \eg, $\ell_{\alpha}\phi \to
\ell_{\beta}\phi$. Contrary to standard leptogenesis, the rates can be large, since the Yukawa couplings $y_{\alpha}$ are not constrained by neutrino masses, and therefore flavour
equilibration is potentially a problem. The question is to what extent it reduces the available parameter space.

There are three different channels contributing to $\Delta L=0$ scatterings: $s$-channel $\ell_{\alpha}\phi \to \ell_{\beta}\phi$, $t$-channel $\ell_{\alpha}\phi^{\dagger}\to
\ell_{\beta}\phi^{\dagger}$, and $t$-channel $\ell_{\alpha}\ell_{\beta}^c\to \phi \phi^{\dagger}$. The reduced cross-sections for these processes can be found in \Ref~\cite{Pilaftsis:2005rv},
and in the limit $M'\ll M$ the total $\Delta L=0$ cross-section is given by
\begin{equation}
\hat{\sigma}_{\alpha \beta}(x)\simeq \frac{5}{4}\frac{|y_{\alpha}|^2 |y_{\beta}|^2}{\pi} \left(\frac{M'}{M}\right)^2x,
\end{equation}
where $x \equiv s/M'^2$. Note that the $N_2$ and $N_3$ contributions are essentially equal. The reaction rate is then obtained using
\begin{equation}
\Gamma^{\Delta L=0}_{\alpha \beta}\equiv \frac{M'z^2}{96\pi^2 \zeta(3)}\int_{x_{\rm thr}}^{\infty} dx \sqrt{x} \mathcal{K}_1(z\sqrt{x})\hat{\sigma}(x),
\end{equation}
where $z\equiv M'/T$ and $\mathcal{K}_1$ is the modified Bessel function of the second kind. This rate, as all the rates entering
the Boltzmann equations, will be compared to the Hubble expansion rate, given by
\begin{equation}
H(z)=1.66 \sqrt{g_{\star}} \frac{M'^2}{z^2 M_{\rm Pl}},
\end{equation}
where $g_{\star}=106.75$ and $M_{\rm Pl}=1.22\times 10^{19}$~GeV.

The Boltzmann equations for leptogenesis are given by 
\begin{eqnarray}
\frac{{\rm d}N_{N_1}}{ {\rm d}z}&=& -D\,(N_{N_1}-N_{N_1}^{\rm eq}),\label{boltz1}\\
\frac{{\rm d}N_{\Delta_{\alpha}}}{ {\rm d}z}&=&\varepsilon_{1\alpha}\,D\,(N_{N_1}-N_{N_1}^{\rm eq})-W^{\rm ID}_{\alpha}\,N_{\Delta_{\alpha}}-
\sum_{\beta\neq \alpha}S^{\Delta L=0}_{\alpha \beta}(N_{\Delta_{\alpha}}-N_{\Delta_{\beta}})\label{boltz2},
\end{eqnarray}
where $N_X$ denotes any particle number or asymmetry $X$ in a portion of comoving volume containing one heavy neutrino in ultra-relativistic
thermal equilibrium, so that $N^{\rm eq}_{N_1}(T\gg M_1)=1$. As a function of $z$, the equilibrium RH neutrino number density is given
by $N_{N_1}^{\rm eq}=0.5\, z^2 \,\mathcal{K}_2(z)$.
Furthermore, we have defined $\Delta_\alpha \equiv \Delta B/3 - \Delta L_\alpha$.
The normalized decay rate is given by $D\equiv \Gamma_D/[H(z)\,z]= K_1 \,z\, \langle{1/\gamma}\rangle$ with the thermally 
averaged dilation factor $\langle{1/\gamma}\rangle$ given by the ratio
of the modified Bessel functions $\mathcal{K}_1(z)/\mathcal{K}_2(z)$. Finally, we defined the normalized inverse decay rate
$W_{\alpha}^{\rm ID}\equiv \Gamma_{\alpha}^{\rm ID}/[H(z)\,z]=0.25\, K_{1\alpha} \,\mathcal{K}_1(z) \,z^3$, and $S^{\Delta L=0}_{\alpha \beta}\equiv 
\Gamma^{\Delta L=0}_{\alpha \beta}/[H(z)\,z]$. Note that the normalized scattering rate is fitted within 10~\% by
\begin{equation}
S^{\Delta L=0}_{\alpha \beta}\simeq 6.5\times 10^{-4}\, |y_{\alpha}|^2 |y_{\beta}|^2\,\left(\frac{M'M_{\rm Pl}}{ M^2}\right)\, z^{-2}.
\end{equation}
In the Boltzmann equations above, spectator processes~\cite{Buchmuller:2001sr,Nardi:2005hs} and the conversion of a 
lepton flavour asymmetry into a $\Delta_{\alpha}$ asymmetry have been neglected, but we have checked that they do
not change our results by more than 20~\%. We have also checked that $\Delta L=1$ scatterings contribute subdominantly in 
the strong washout regime under consideration. As for $\Delta L=2$ processes their rates are suppressed by the small neutrino masses and can be safely neglected.

\begin{figure}[t]
\includegraphics[width=0.7\textwidth]{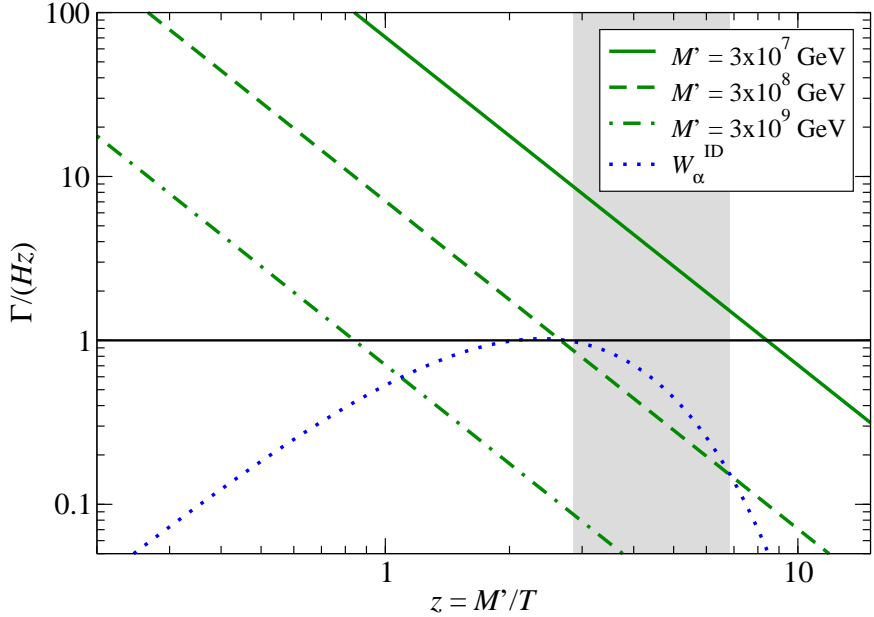}
\caption{Scattering rate $S^{\Delta L=0}_{\alpha \beta}$ leading to flavour equilibration vs. $z=M'/T$ for different values 
of the lightest heavy neutrino mass $M'$. 
Also plotted is the inverse decay rate $W_{\alpha}^{\rm ID}$ in flavour $\alpha$, with $K_{1\alpha}=3.5$. The mass
hierarchy has been fixed to $M/M'=10$. The shaded region denotes the time at which the asymmetry is produced (see text).} \label{fig:1}
\end{figure}

We show in \fig~\ref{fig:1} how the scattering rate normalized to the Hubble expansion rate varies with $M'$, for a fixed hierarchy $M/M'=10$ and
washout $K_{1\alpha}=3.5$ , and for Yukawa couplings such that
the baryon asymmetry of the Universe is produced [\cf~Eq.~(\ref{finalasym2})]. The shaded regions denote the asymmetry production time,
roughly when $z_B-2<z<z_B+2$ with $z_B(K)\simeq 2+4 K^{0.13}\exp(-2.5/K)\simeq 5$~\cite{Blanchet:2006dq}. Note that even though the flavour
equilibrating scattering rate falls out of equilibrium before entering the shaded region for $M'\gtrsim 3\times 10^8$~GeV, there is 
still a residual effect, which suppresses the asymmetry by a factor of 2--3. The reason is that the scattering rate is within a factor of 
2 of the inverse decay rate during all the asymmetry production, and therefore some flavour equilibration is achieved.

\begin{figure}[t]
\includegraphics[width=0.7\textwidth]{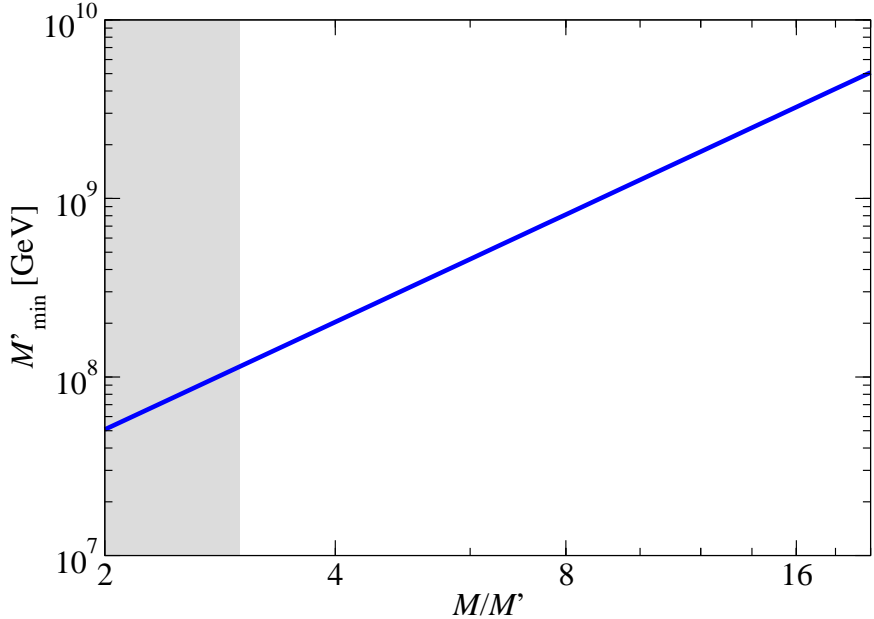}
\caption{Lower bound on the lightest heavy neutrino mass, $M'_{\rm min}$, for successful leptogenesis, vs. the hierarchy parameter $M/M'$. 
The washout parameter is fixed to $K_{1\alpha}=3.5$,
and the asymmetry is mainly produced in flavour $\alpha$.} \label{fig:2}
\end{figure}

Since $\Gamma^{\Delta L=0}/H \propto 1/M'$ for a fixed mass hierarchy $M'/M$, it is clear that the scale of leptogenesis cannot be arbitrarily low. We have 
solved numerically the system of Boltzmann equations~(\ref{boltz1}) and (\ref{boltz2}) without neglecting the contribution
from any flavour, so that
\begin{equation}
\eta_{B}=0.88\times 10^{-2} \sum_{\alpha}\varepsilon_{1,\alpha} \;\kappa(K_{1\alpha}),
\end{equation}
and we have required that the final asymmetry should fall within the 3$\sigma$ range of the observed baryon asymmetry, \ie, $\eta_B>5.75\times 10^{-10}$.
We then calculated the lower bound for successful leptogenesis for different hierarchies
$M/M'$, and the result is shown in Fig.~\ref{fig:2}.
We find that the lowest bound on the scale of leptogenesis is given by
\begin{equation}
M'_{\rm min}\gtrsim 10^{8}~{\rm GeV},
\end{equation}
obtained when $M/M' =3$. Increasing the hierarchy increases the scattering rate for a fixed value of the baryon asymmetry and thus the lower bound increases. Note that the shaded area here
denotes the non-hierarchical region $M<3\, M'$, where our results are not valid. It should be noted that this lower bound was obtained
 maximizing the asymmetry with respect to $K_1$ and $K_{1\alpha}$ and the phases contributing to the $C\!P$ asymmetry. We have chosen $K_{1\alpha}=3.5$, the smallest value in the strong washout
 regime~\cite{Blanchet:2006be}. Moreover, we have fixed $K_{1\beta\neq \alpha}=11$, so that $K_1=25.5$, which maximizes the asymmetry.\footnote{If 
 $|Y_{1\beta}|$ is too close to $|Y_{1\alpha}|$, the washout in all flavours are similar, and the
asymmetry is suppressed since $\varepsilon_1\simeq 0$. However, if $|Y_{1\beta}|$ is much bigger than $|Y_{1\alpha}|$, then the $C\!P$ asymmetry is suppressed
[\cf~Eq.~(\ref{Eq:AssFromd=6})].}

In our model leptogenesis can take place more than one order of magnitude below the usual lower bound, given by $3\times 10^9$~GeV at the onset of the strong washout. Unfortunately, it
turns out that the coefficient of the $d=6$ operator leading to non-unitarity in \eq~(\ref{Eq:EpsAndNonU}) is expected to be very suppressed, of order $\eta_{\alpha \beta} \sim
10^{-16}$. The reason is mainly the high scale of $M'$, and thus $M$, needed for successful leptogenesis. Moreover, the case $M/M'=3$ provides the largest non-unitarity possible, since the
latter decreases when the scale of $M$ increases. For instance, with $M/M'=10$, the Yukawa coupling needed for leptogenesis is $y\simeq 0.07$, in which case we find the lower bound $M'_{\rm
min}\simeq 10^9$~GeV and the expected deviation from unitarity is $\eta_{\alpha \beta} \sim 10^{-18}$. We conclude that, in the considered scenario, successful leptogenesis is incompatible with observable non-unitarity signals.

\section{Summary and discussion}
\label{sec:conclusions} We have investigated the relation between non-unitarity of the leptonic mixing matrix and baryogenesis via thermal leptogenesis. We have first studied how all
parameters of the canonical \type{I} seesaw mechanism can, in principle, be reconstructed from the neutrino mass matrix and a measurement of the deviation of the effective low-energy leptonic
mixing matrix from unitary. In the effective low-energy theory, neutrino masses and non-unitarity are encoded in the lepton number violating $d=5$ (Weinberg) operator and in the
$d=6$ operator contributing to the neutrino kinetic terms after electroweak symmetry breaking, respectively.

For the case that the mass $M'$ of the lightest right-handed neutrino is lighter than the masses of the others, we show that its decay asymmetries for flavour-dependent leptogenesis can be
expressed in terms of two contributions, one depending on the unique $d=5$ operator generating neutrino masses and one depending on the $d=6$ operator associated with
non-unitarity. We have argued that in low-energy seesaw scenarios, where small lepton number violation explains the smallness of neutrino masses, the lepton number conserving $d=6$
operator contribution, linked to non-unitarity, generically dominates over the $d=5$ operator contribution which results in a strong enhancement of the flavour-dependent decay asymmetries
without any resonance effects. We have referred to this case as \emph{non-unitarity driven leptogenesis}.

To calculate the produced final baryon asymmetry, however, we found that lepton flavour equilibrating effects directly related to non-unitarity play a crucial role and their effects have to
be included. In the simple realization of non-unitarity driven leptogenesis considered here, they turn out to forbid lowering the leptogenesis scale down to the TeV scale. Nevertheless, lowering of the leptogenesis scale, \ie, the mass of the lightest right-handed
neutrino $M'$, to about $10^8$~GeV is possible, which is more than one order of magnitude below the scale of standard thermal
leptogenesis. The reduced leptogenesis scale in non-unitarity driven leptogenesis can improve consistency between leptogenesis and gravitino (or similar) constraints in
supergravity theories. On the other hand, the deviation from unitarity for the case of $M' \gtrsim 10^8$~GeV is far below the experimentally accessible region.

\begin{acknowledgments}
We would like to thank Jacobo Lopez Pavon for useful discussions. S.B. would like to thank the Max-Planck-Institut for Physics for hospitality when this work started. This work was supported
by the European Community through the European Commision Marie Curie Actions Framework Programme 7 Intra-European Fellowship: Neutrino Evolution [M.B.]. S.A., M.B. and E.F.M. acknowledge
support by the DFG cluster of excellence ``Origin and Structure of the Universe''. S.A. and E.F.M. also acknowledge support from the European Community under the European Commission
Framework Programme 7 Design Study: EUROnu, Project Number 212372.
\end{acknowledgments}

\end{document}